\documentclass[twocolumn,amsmath,prl]{revtex4}
\usepackage{graphicx,amsmath,bm}
\newcommand{\avg}[1]{\langle #1 \rangle}
\begin{document}
\title{Oscillations of 2D Rashba system spin polarization
in quantizing magnetic field}
\author{I. I. Lyapilin}\email{Lyapilin@imp.uran.ru}
\author{A. E. Patrakov}
\affiliation{Institute of Metal
Physics, UD of RAS, Yekaterinburg, Russia}
\maketitle
\section{Introduction}

There are two main types of spin-orbit coupling in quantum wells based on
semiconductors having the Zincblende structure: Rashba interaction
\cite{rashba}, caused by the structural asymmetry of the quantum well, and
Dresselhaus interaction \cite{dressel}, originating due to the structural
inversion asymmetry of the bulk material.
In the approximation linear in the electron momentum, both contributions can
be formally represented in ways similar to each other.
The spin-orbit interaction (SOI) leads to correlation of the translational
and spin motion of electrons.
Such correlation is the origin of many transport phenomena observed in such
systems.
Among such phenomena, there are beats in Shubnikov---de Haas oscillations
\cite{Das}, spin accumulation \cite{Edel}, magneto-electric effect
\cite{Levitov}, etc.
In the cases when the SOI is in some sense small, one can perform a
canonical transformation of the Hamiltonian, that decouples kinetic and spin
degrees of freedom.
All other terms in the Hamiltonian, describing the interaction
of electrons with the lattice and external fields (if any) also undergo the
transformation.
In the latter case, the effective interaction of electrons in the system
with external fields appears, which leads to resonant absorption of the
field energy not only at the frequency of the paramagnetic resonance $\omega_s$
or cyclotron resonance $\omega_c$, but also at their linear combinations
\cite{Rash}.
A gauge-invariant theory of this phenomenon has been developed in
\cite{Kalash}.
This interaction also determines the gauge-invariant equations of motion for
macroscopic variables characterizing the system.
We apply the method developed in \cite{Kalash} in order to describe the
kinetics of 2D conductivity electrons in electric and quantizing magnetic
fields when the spin-orbit interaction is present.
We write down and solve the macroscopic equations for the energy balance of
the spin subsystem and for transverse spin components; consider the
magneto-electric effect and find the polarization of spins of the charge
carriers in quantizing magnetic field.
We show that, in quantizing magnetic field, the oscillations of the spin
polarization become ``anomalously'' large.

\section{Effective interaction}
The Hamiltonian of the 2D system can be written as:
\begin{equation}\label{1}
\mathcal H = H_k + H_s + H_{ks} + H_{ef} + H_v + H_{ev}.
\end{equation}
Here $H_k$ and $H_s$ are kinetic and Zeeman energies, respectively, in
the magnetic field $\bm H = (0, 0, H)$.
$H_{ef}$ is the Hamiltonian of the electrons' interaction with the
electric field.
$H_{ev}$ and $H_v$ are Hamiltonians of
the electron-lattice interaction and of the lattice itself, respectively.
$H_{ks}(p)$ is the interaction between translational and spin degrees of
freedom:
\begin{eqnarray}\label{2}
H_k & = & \sum_i\frac{(\bm p_i-(e/c)\bm A (\bm x_i))^2}{2m},\nonumber\\
H_s & = & \hbar\omega_s \sum S_i^z,\quad
\hbar\omega_s=g \mu_0 H.
\end{eqnarray}
$S_i^\alpha$ and $p_i^\alpha$ are operators of the components of the spin
and kinetic momentum of the $i$th electron.
$g$ denotes the gyromagnetic ratio of electrons, and $\mu_0$ is Bohr
magneton.
\begin{equation}\label{3}
H_{ef} = -e\bm E \sum_i\bm r_i.
\end{equation}
The most general form of $H_{ks}(\bm p)$ is:
\begin{eqnarray}\label{4}
H_{ks}(\bm p) = \sum_j \bm f(\bm p_j) \bm S_j =
\sum_{\alpha_1\alpha_2\ldots\alpha_s;\beta} \phi^{\alpha_1\alpha_2\ldots\alpha_s;\beta},\\
\phi^{\alpha_1\alpha_2\ldots\alpha_s;\beta} = const
\sum_j p_j^{\alpha_1} p_j^{\alpha_2} \ldots p_j^{\alpha_s} S_j^\beta
\nonumber
\end{eqnarray}
Here $\bm f(\bm p_j)$ is a pseudo-vector, components of which can be
represented as a form of order $s$ in the components of the kinetic momentum
$p_j^\alpha$.

Now we perform the canonical transformation of the Hamiltonian.
Up to the terms linear in $T(t)$, we have:
\begin{equation}\label{5}
\tilde H = e^{T(p)}\mathcal H e^{-T(p)}\approx
\mathcal H +[T(p), \mathcal H].
\end{equation}
The operator of the canonical transformation $T(p)$ has to be determined
from the requirement that, after the transformation, the $k$ and $s$
subsystems become independent.
This requirement can be written as the following condition:
\begin{equation}\label{6}
H_{ks}(p) + [T(p), H_k + H_s]=0.
\end{equation}

Note that, after the canonical transformation, the $H_k$ and $H_s$ operators
are the integrals of motion if $H_{ev}=0$.
We assume the specific form of the $H_{ks}$ term, namely, Rashba interaction,
which is non-zero even in the linear order in momentum:
\begin{equation}\label{7}
H_{ks}(p) =
\alpha \varepsilon_{zik} \sum_j S^i_j p^k_j =
\frac{i\alpha}{2} \sum_j (S^+_j p^-_j - S^-_j p^+_j),
\end{equation}
$$
S^\pm = S^x \pm i S^y, \quad p^\pm = p^x \pm i p^y.
$$
Here $\alpha$ is the constant characterizing the spin-orbit interaction,
$\varepsilon$ is the fully-antisymmetric Levi---Chivita tensor.

Inserting the operator (\ref{7}) into the general solution for Eq. (\ref{6}),
we obtain:
\begin{equation}\label{8}
T(p)=\frac{i \alpha}{2\hbar(\omega_c-\omega_s)}
\sum_j (S^+_j p^-_j - S^-_j p^+_j).
\end{equation}
The criteria for the applicability of this theory is that, for
characteristic values of the electron momentum $\bar p$, the inequality
$\alpha \bar p \ll \hbar(\omega_c-\omega_s)$ should hold.

Using the explicit form of the operator $T(p)$, we have for the
effective interaction:
\begin{eqnarray}\label{9}
eE^\alpha(t) [x_j^\alpha, T(p)]&=&
-\frac{e\alpha}{2\hbar(\omega_c-\omega_s)}
(S^+ E^-(t) + S^- E^+(t)),\nonumber \\
S^\alpha&=&\sum_i S^\alpha_i.
\end{eqnarray}
The renormalized interaction with the electric field contains only spin
variables, thus, it affects only spin evolution of the conductivity
electrons.

Now, taking into account the explicit form of the operator that defines the
canonical transformation, we find the operators of power absorbed by the
electron subsystem:
\begin{multline}\label{10}
\dot H_{e(f)}(t)=\dot H _{k(f)}(t) + \dot H_{s(f)}(t)
=\frac{e \bm E(t) \bm p}{m} +\\+\frac{i e\alpha \omega_s}
{2(\omega_c-\omega_s)}\{S^-E^+(t) - S^+E^-(t)\}
=J_e^\alpha E^\alpha(t).
\end{multline}
Here
\begin{eqnarray}\label{11}
J_e^\alpha &=& e V^\alpha,\nonumber\\
V^\alpha &=& \frac{p^\alpha}{m}+
\frac{1}{i\hbar}[x^\alpha,H_{ks}(p)] + \frac{1}{m}[T(p),p^\alpha].
\end{eqnarray}
The operator $V^\alpha$ is the transformed electron velocity in the zeroth
order in the field.
The expression $J_e^\alpha E^\alpha(t)$ is the operator of power absorbed by
both translational and spin degrees of freedom due to the interaction of the
conductivity electrons with the electric field.
One can write $V^\alpha = V^\alpha_k + V^\alpha_s$, where
\begin{equation}\label{12}
V^\pm_k = \frac{p^\pm}{m},\quad
V_s^\pm = \mp\frac{i \alpha}{2\hbar(\omega_c-\omega_s)}S^\pm.
\end{equation}
\section{Balance equations}
Bearing the evolution of the spin subsystem in mind, we write down
the balance equations for the Zeeman energy and the transverse
components of spin:
\begin{multline}\label{13}
\dot{H_s} = \frac{\alpha e \omega_s}{2 i \hbar(\omega_c-\omega_s)}
\{S^+ E^-(t) - S^- E^+(t)\} +\\+ \frac{1}{i\hbar}[H_s,\tilde{H}_{ev}]
\end{multline}
\begin{equation}\label{14}
\dot{S}^{\pm} = \mp i\omega_s S^\pm \mp
\frac{i e \alpha}{\hbar(\omega_c-\omega_s)} S^z E^\pm(t) +
\frac{1}{i\hbar}[S^\pm, \tilde{H}_{ev}].
\end{equation}
We describe the state of the non-equilibrium system with the
average values of the following operators: $H_k$, $H_s$, $N$,
$H_v$ ($N$ is the operator for number of electrons). In this case,
for the non-equilibrium statistical operator (NSO) $\rho(t)$
\cite{Kalash}, we have:
\begin{widetext}
\begin{equation}\label{15}
\rho(t,0) = \rho_q(t,0)-\int_{-\infty}^0 dt_1 e^{\varepsilon t_1}
\{\int_0^1 d\tau\rho_q^\tau\dot{S}(t+t_1,t_1) \rho_q^{(1-\tau)}
+ i L_{ef}(t+t_1)\rho_q(t+t_1,t_1)\},
\end{equation}
$$
A(t,t_1) = e^{it_1L}A(t,0),\qquad
iL_{ef}A=(i\hbar)^{-1}[A, H_{ef}].
$$
\end{widetext}
Here $S(t)$ is the entropy operator:
\begin{equation}\label{16}
S(t) = \Phi(t) + \beta_k (H_k-\mu N)
\beta_s H_s + \beta(H_v+\tilde H_{ev}).
\end{equation}
The parameters $\beta_k$, $\beta_s$, $\beta$ have the meaning of the inverse
effective temperatures of the kinetic and spin electron subsystems and the
inverse equilibrium temperature of the lattice.
\begin{eqnarray}\label{17}
\dot S(t,0) = \frac{\partial S(t,0)}{\partial t}
+\frac{1}{i\hbar} [S(t,0),\tilde H_{ev}]
\end{eqnarray}
is the entropy production operator.
$\rho_q(t,0)=exp\{-S(t)\}$ is the quasi-equilibrium statistical operator.

Averaging the operator equations of motions for the spin subsystem yields:
\begin{multline}\label{19}
\partial_t \avg{H_s} = \frac{\alpha e\omega_s}{2i\hbar(\omega_c-\omega_s)}
\{\avg{S^+} E^-(t) - \avg{S^-} E^+(t)\} +\\+\avg{\dot{H}_{s(v)}},
\end{multline}
\begin{multline}\label{20}
\partial_t \avg{S^{\pm}} = \mp i\omega_s \avg{S^\pm} \mp
\frac{ie\alpha}{\hbar(\omega_c-\omega_s)} \avg{S^z} E^\pm(t) +\\+
\avg{\dot S^\pm_{(v)}}.
\end{multline}

The first term in the right hand side of Eq. (\ref{19}) represents Joule
heat acquired by the spin subsystem of conductivity electrons. The second
term describes the relaxation of the longitudinal component of electrons'
spin.

Let's turn our mind to the equation (\ref{20}) describing the evolution of
the transverse components of the spin. The collisional term in that formula
has the order of $\avg{S^\pm}\cdot\nu_2$, where $\nu_2$ is the relaxation
frequency of the transverse spin. Below we shall obtain an explicit
expression for $\nu_2$. Now the balance equation for the transverse spin is
easily solved, and for the steady-state case we obtain the expression that
determines the polarization of electron spins $m^{\pm}=g\mu\avg{S^\pm}$:
\begin{equation}\label{21}
m^{\pm} = \pm \frac{i\alpha g\mu e E^\pm}
{\hbar(\omega_c-\omega_s)}
\cdot\frac{\avg{S^z}}{\mp i(\omega-\omega_s) + \nu_2}.
\end{equation}
\begin{equation}\label{22}
\avg{S^z} = (1/2)\sum_\mu\{f_{\mu\uparrow}-f_{\mu\downarrow}\},
\end{equation}
where the symbols $\uparrow$, $\downarrow$ denote the spin orientation with
respect to the $z$ axis.

It follows from Eqs. (\ref{21}), (\ref{22}) that the dependence of the
average magnetic moment of electrons $m^{\pm}$ upon the magnetic field is
determined by the form of the density of states $\rho(\varepsilon)$.
Within the framework of the self-consistent Born approximation, we have
\cite{Ando}:
\begin{equation}\label{23}
\rho(\varepsilon) = \rho_0 [ 1 + 2 \sum\limits_{l=1}^\infty
e^{(-\pi l/\omega_c\tau_0)} \cos(\frac{2\pi l \varepsilon}{\hbar \omega_c})].
\end{equation}
Here $\rho_0=m/\hbar^2\pi$ is the density of states in zero magnetic field,
$\tau_0$ is the relaxation time in zero magnetic field.
We have omitted the Zeeman splitting from the expression for the density of
states.
It follows from Eq. (\ref{23}) that, when $\omega_c\tau_0 \ll 1$, the
density of states is $\rho(\varepsilon)=\rho_0$.
When the magnetic field is
increased, so that $\omega_c\tau\leq 1$, it is sufficient to leave only the
first term in the sum.
In that case, the oscillating expression for the density of states has the
sinusoidal form.

Results of the numerical calculation of the magnetization $m^+$ are
presented on Fig. 1. They were obtained for the following parameters:
$m = 0.067m_0$ ($m_0$ is the free electron mass),
the Fermi energy $\mathcal E_F = 10$~meV,
the 2D electron mobility $\mu\approx 0.1-1.0\times 10^7$~cm$^2$/Vs.
It follows from the numerical analysis that the dependence of the electron
magnetization upon the magnetic field has oscillating character, and the
amplitude of the oscillations is very sensitive to temperature.

Now consider the power absorbed by the spin subsystem.
From Eq. (\ref{19}), we have:
\begin{equation}\label{24}
Q=\beta_s\left(\frac{\alpha e \omega_s E_\perp}
{2\hbar(\omega_c- \omega_s)}\right)^2
2\pi\{G^{+-}(\omega)+G^{-+}(-\omega)\},
\end{equation}
where $G^{\pm\mp}(t)$ are Green's functions:
\begin{eqnarray}\label{25}
G^{\pm\mp}(t) &=& \theta(-t) e^{\varepsilon t}(S^\pm,S^\mp(t)) =
\int_{-\infty}^\infty d\omega e^{i\omega t}G^{\pm\mp}(\omega),\nonumber\\
(A,B)&=&\int_0^1 d\tau \operatorname{Sp} (A \rho_q^\tau
\Delta B \rho_q^{(1-\tau)}), \nonumber\\
\Delta A &=& A-\operatorname{Sp}\{A\rho_q\}.
\end{eqnarray}
$\theta(t)$ is the unit step (Heaviside) function.

Making a chain of motion equations for the Green's function and neglecting
terms of more than second order in the interaction $H_{ev}$, and keeping
only zeroth order in thermodynamic forces in terms
having the first and the second order in $H_{ev}$, we obtain:
\begin{equation}\label{25a}
G^{+-}(\omega) =
\frac{1}{\pi\beta\hbar\omega_s}
\frac{\avg{S^z}}{i\omega_s + M^{+-}(\omega)}.
\end{equation}
The quantity $M^{+-}(\omega)$ is the mass operator calculated in the second
order in the electron-lattice interaction and in zeroth order in
thermodynamic forces:
\begin{equation}\label{26}
M^{+-}(\omega) =
\frac{\beta\hbar\omega_s}{\avg{S^z}}\int_{-\infty}^0 dt
e^{\epsilon t}(\dot S^+_{(v)},\dot S^-_{(v)}(t))
\end{equation}

The imaginary part of the mass operator
$\operatorname{Im}M^{+-}(\omega)=\delta\omega_s$ determines the frequency
shift for electron spins, while its real part
$\operatorname{Re}M^{+-}(\omega)=\nu_2(\omega)$ plays the role of the
inverse relaxation time for the transverse spin components.

The canonical transformation that we performed earlier for decoupling the
kinetic and spin subsystems also leads to the renormalization of the
electron-lattice interaction Hamiltonian.
The effective Hamiltonian of the electron-lattice interaction now becomes
$H_{el}+[T, H_{el}]$.
Note that the Hamiltonian $H_{el}$ can be presented in the form:
$$
H_{el}= H_{el}'+H_{el}'',
$$
where $H'_{el}$ is the Hamiltonian of the spin-independent part of the
electron-lattice interaction (it is responsible, e.g., for the electron
momentum relaxation), and $H_{el}''$ is the spin-dependent part, responsible
for the relaxation of the electron magnetization.
Since the canonical transformation operator $T(p)$ depends upon the electron
spin, the Hamiltonian of the total spin-lattice interaction acquires
the form:
$$
H''_{el}+[T(p), H_{el}'].
$$
Here we neglected higher-order terms in the spin-orbit interaction, that
arise from the $[T(p), H_{el}']$ commutator.
It can be shown that situations are possible where the contribution of the
renormalized part of the conductivity electrons' interaction with the
lattice has the same order as the contribution from the ordinary
electron-phonon interaction, but those two contributions have, obviously,
substantially different dependencies upon the temperature and the magnetic
field.

\begin{figure}
\center\includegraphics[width=8cm]{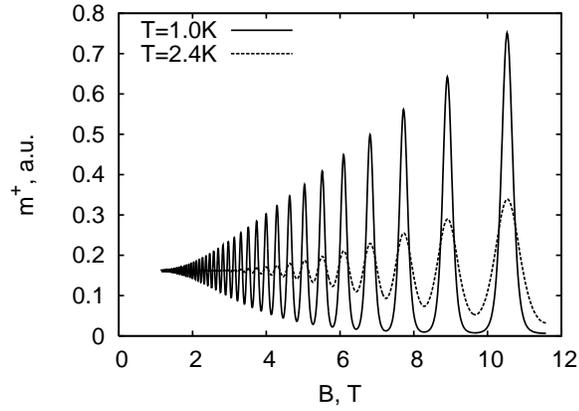} \caption{The
magnetiztion of 2D electron gas vs the magnetic field for
different temperatures}
\end{figure}

\end{document}